\documentclass[nofootinbib,aps,preprint,superscriptaddress]{revtex4}
\usepackage{graphicx,amssymb,amsmath}

\newcommand{\ben}{\begin{enumerate}}
\newcommand{\een}{\end{enumerate}}

\newcommand{\bea}{\begin{eqnarray}}
\newcommand{\eea}{\end{eqnarray}}
\newcommand{\beqa}{\begin{eqnarray}}
\newcommand{\eeqa}{\end{eqnarray}}
\newcommand{\beq}{\begin{equation}}
\newcommand{\eeq}{\end{equation}}
\newcommand{\bay}{\begin{array}}
\newcommand{\eay}{\end{array}}

\def\gsim{\ \rlap{\raise 3pt \hbox{$>$}}{\lower 3pt \hbox{$\sim$}}\ }
\def\lsim{\ \rlap{\raise 3pt \hbox{$<$}}{\lower 3pt \hbox{$\sim$}}\ }

\def\Ord{{\cal O}}
\def\lt{\left}
\def\rt{\right}

\def\Im{{\cal I}m}
\def\Re{{\cal R}e}
\def\qp2{\lt|\frac{q}{p}\rt|^2}
\def\pq2{\lt|\frac{p}{q}\rt|^2}
\def\Af2{\lt|A_{Bf}\rt|^2}
\def\Abf2{\lt|\bar A_{Bf}\rt|^2}
\def\Afb2{\lt|A_{B\bar f}\rt|^2}
\def\Abfb2{\lt|\bar A_{B\bar f}\rt|^2}
\def\ss2{s_{12}}
\def\ss3{s_{13}}
\def\eg{{\it e.g.}}

\begin{document}

\preprint{\vbox
{\hbox{}
\hbox{}
\hbox{}
\hbox{hep-ph/0702011}}}

\vspace*{3cm}

\title{Enhanced effects on extracting $\gamma$ from untagged
$B^0$ and $B_s$ decays}

\def\addtech{Department of Physics,
Technion--Israel Institute of Technology,\\
Technion City, Haifa 32000, Israel\vspace*{6pt}}

\def\addLjub{Department of Physics,
University of Ljubljana,
Jadranska 19, \\
1000 Ljubljana, Slovenia\vspace*{20pt}}

\def\addstefan{J. Stefan Institute,
Jamova 39, P.O. Box 3000, 1001 Ljubljana, Slovenia\vspace*{6pt}}

\author{Michael Gronau}\affiliation{\addtech}
\author{Yuval Grossman}\affiliation{\addtech}
\author{Ze'ev Surujon}\affiliation{\addtech}
\author{Jure Zupan}\affiliation{\addstefan}
\affiliation{\addLjub}

\begin{abstract}
The weak phase $\gamma$ can be determined using untagged $B^0\to DK_S$
or $B_s\to D\phi, D\eta^{(')}$ decays. In the past, the small lifetime
difference $y\equiv \Delta\Gamma/(2\Gamma)$ has been neglected in $B^0$,
while the CP violating parameter $\epsilon\equiv 1-|q/p|^2$ has been
neglected in both $B^0$-$\bar B^0$ and $B_s$-$\bar B_s$ mixing.
We estimate the effect of neglecting $y$ and $\epsilon$. We find that
in $D$ decays to flavor states this introduces a systematic error, which is enhanced 
by a large ratio of Cabibbo-allowed to doubly Cabibbo-suppressed $D$ decay amplitudes.
\end{abstract}
\maketitle


\section{Introduction}
\label{time-indep}
\setcounter{equation}{0}

The interference between the decay chains
\mbox{$B\to DK\to f_DK$} and \mbox{$B\to \bar DK\to f_DK$} is
commonly utilized in methods for extracting the phase $\gamma={\rm
arg}\lt(-V_{ud}V_{ub}^*V_{cb}V_{cd}^*\rt)$.
The original idea, applied to these processes and to $B_s\to D\phi$,
was put forward in~\cite{GW,GL}.
Several variants of the idea can be found
in~\cite{ADS,GLS,GGSZ,AS,ID,multi,AS2}.  The determination of $\gamma$
based on these methods is theoretically very clean.
The uncertainty in the value of $\gamma$ is mainly due to low statistics
of the relevant processes.
Time-dependent decays $B^0\to DK_S$ studied in~\cite{Kayser:1999bu}
require flavor tagging of the initial $B^0$, for which one pays a
price in statistics.
This is also the case for time-dependent $B_s\to D\phi$.
Ref.~\cite{Dunietz:1995cp} discussed the extraction of $\gamma$
in untagged $B_s$ decays, taking advantage of the finite nonzero
width-difference between the two $B_s$ mass-eigenstates.
Recently it was shown that $\gamma$ can also be determined using
untagged non-strange neutral $B$ decays~\cite{GGSSZ}, where the width-difference
is too small to be measured.
In this method, combining charged along with neutral $B$ decays, and
using several $D$ decay modes, reduces the statistical error on the
value of $\gamma$.

The two parameters describing the width-difference and CP violation in 
mixing in the $B^0$ systems, 
\beq
   y\equiv \frac{\Delta\Gamma}{2\Gamma}~,\quad
   \epsilon\equiv 1 -\left|\frac{q}{p}\right|^2~, \label{eq:defs}
\eeq
are estimated to be below a level of $10^{-2}$~\cite{Bigi:1987in,Lenz:2004nx}.
(Note that $\epsilon$ is not a standard notation).
Thus, usually these parameters can be safely neglected when
dealing with untagged $B^0\to DK_S$. Indeed, both $y$ and $\epsilon$ were
neglected in \cite{GGSSZ}.

In this work we consider the extraction of $\gamma$ from both
untagged $B_d\to DK_S$ (here $B_d \equiv B^0$)
and untagged $B_s\to D\phi$, keeping explicitly $y$ and $\epsilon$.
We find that in $D$ decays to flavor states,
the sensitivity of $\gamma$ to $y$ and $\epsilon$ is enhanced by
ratios of Cabibbo-allowed and doubly Cabibbo-suppressed $D$ decay
amplitudes.  In such cases, neglecting $y$ and $\epsilon$ might result in a relatively
large systematic error.  This effect is expected to be more important
in $B_s$ decays,
as the width-difference in the $B_s$ system is expected to be much larger than
in $B_d$~\cite{Lenz:2004nx,Lenz:2006hd}. The effect in $B_s$ decays may be
taken into account through a measurement of the $B_s$ width-difference.
Eventually, when very large  data sets will be available, a most precise determination
of $\gamma$ will require using untagged $B_d$ and $B_s$ decays  together with
charged $B$ decays. The purpose of this note is to point out the nontrivial
enhancement of the effect of nonzero $y$ and $\epsilon$, which may become relevant
at that point.

\section{Two-body $D$ decays}
\label{twobody}

Our analysis applies to both $B_d\to DK_S$ and $B_s\to D\phi, D\eta^{(')}$.
Working along the lines of \cite{GGSSZ}, we define for the first process,
\beq\label{An}
   A(B^0\to \bar{D}^0 K_S) = A_n~, \qquad
   A(B^0\to D^0 K_S) = A_n r_n e^{i\lt(\delta_n+\gamma\rt)}~,
\eeq
where we take $A_n$ and $r_n$ to be real and positive,
$0\le\delta_n<2\pi$ and $\gamma$ are the strong and weak phases,
respectively. Neglecting $CP$ violation
in $K^0$-$\bar K^0$ mixing, amplitudes for $CP$-conjugate decays are given 
by inverting the sign of the weak phase $\gamma$.

The parameter $r_n$ is expected to be 
around $0.4$ since it contains a Cabibbo-Kobayashi-Maskawa (CKM) factor
$|V^*_{ub}V_{cs}/V^*_{cb}V_{us}|\simeq 0.4$. It is not expected to be much
smaller than 
this value unless color-suppression in $\bar b\to \bar u c \bar s$ is more
effective than in $\bar b \to \bar c u\bar s$.
An estimate of this effect can be obtained by comparing the ratios 
$r_n^{*}$ and $r_c^{*}$ measured in $B^0\to D K^{*0}$ and $B^+\to D K^{*+}$ decays. 
[The parameter $r^*_n$ is defined in analogy with (\ref{An}), while $r^*_c$ is the 
corresponding parameter in charged $B$ decays.]
A recent BaBar analysis found an upper bound $r_n^{*}<0.4$ at 
$90\%$~CL \cite{Aubert:2006qn},
while a value $r_c^*=0.11^{+0.08}_{-0.11}$ was obtained by averaging a few 
measurements by BaBar and Belle~\cite{tJampens}. Using isospin and neglecting 
a pure annihilation amplitude describing $B^+\to D^+ K^{*0}$, one has \cite{GGSSZ}
\beq
r_n^{*}=\sqrt{\frac{\Gamma(B^+\to \bar D^0 K^{*+})}{\Gamma(B^0\to \bar D^0 K^{*0})}}r_c^*=(4.0\pm0.5)r_c^*=0.4^{+0.3}_{-0.4}~.
\eeq
Similar notations and considerations apply to $B_s\to D\phi, D\eta^{(')}$.

Considering $D$ decays, we define amplitudes into a
(possibly quasi) two-body final state $f_D$ and its CP-conjugate state $\bar
f_D$ by
\beq
   A(\bar{D^0}\to f_D) =
   A(D^0\to \bar f_D) \equiv A_f~, \qquad
   A(D^0\to f_D) =
   A(\bar{D^0}\to \bar f_D) \equiv A_f r_f e^{i\delta_f}~.
\eeq
By convention $A_f\ge 0, r_f \ge 0, 0\le \delta_f < \pi$.
We neglect CP violation related to $D$ mesons,
as well as $D^0$-$\bar D^0$ mixing effects, which have been shown to be very
small~\cite{GSZ}. The parameters $r_f, \delta_f$ depend on the final state
$f_D$. While $r_f=1$ for CP-eigenstate (\eg~$f_D =K^+K^-$) and $r_f = \Ord(1)$
for non-CP flavorless states (\eg~$f_D=K^{*+}K^-$), $r_f$ is very small for flavor states
(\eg~$f_D = K^+\pi^-$), $r_f \simeq 0.06 \sim \tan^2\theta_c$~\cite{pdg}. In this note
we focus on a particular effect which applies to the latter class of final states.

The amplitudes of the four decay chains are given by
\beqa
   A_{Bf} &\equiv& A(B^0\to f_DK_S) =
   A_nA_f\lt[r_ne^{i\lt(\delta_n+\gamma\rt)}r_fe^{i\delta_f}
   +1\rt]~, \nonumber\\
   \bar A_{Bf} &\equiv& A\lt(\bar B^0\to f_DK_S\rt) =
   A_nA_f\lt[r_fe^{i\delta_f}
   +r_ne^{i\lt(\delta_n-\gamma\rt)}\rt]~,  \nonumber\\
   A_{B\bar f} &\equiv& A\lt(B^0\to \bar f_DK_S\rt) =
   A_nA_f\lt[r_fe^{i\delta_f}
   +r_ne^{i\lt(\delta_n+\gamma\rt)}\rt]~, \nonumber\\
   \bar A_{B\bar f} &\equiv& A\lt(\bar B^0\to \bar f_DK_S\rt) =
   A_nA_f\lt[r_ne^{i\lt(\delta_n-\gamma\rt)}r_fe^{i\delta_f}
   +1\rt]~,   \label{eq:Abf}
\eeqa
with magnitudes
\beqa
   \lt|A_{Bf}\rt| &=&
   A_nA_f\sqrt{1+r_n^2 r_f^2+2r_n r_f
   \cos\lt(\delta_f+\delta_n+\gamma\rt)}~,  \nonumber\\
   \lt|\bar A_{Bf}\rt| &=&
   A_nA_f\sqrt{r_n^2+r_f^2+2r_n r_f
   \cos\lt(\delta_f-\delta_n+\gamma\rt)}~,   \nonumber\\
   \lt| A_{B\bar f}\rt| &=&
   A_nA_f\sqrt{r_n^2+r_f^2+2r_n r_f
   \cos\lt(\delta_f-\delta_n-\gamma\rt)}~,  \nonumber \\
   \lt|\bar A_{B\bar f}\rt| &=&
   A_nA_f\sqrt{1+r_n^2 r_f^2+2r_n r_f
   \cos\lt(\delta_f+\delta_n-\gamma\rt)}~.
   \label{eq:abs}
\eeqa

We note that the phase of $\bar A_{Bf}/A_{Bf}$ depends also
 on $\gamma$. This dependence does not disappear in the limit $r_f \to 0$, in
 contrast to the dependence of the magnitudes of the amplitudes on $\gamma$.
 This will turn out to be a crucial point in our argument in the
 next section, when discussing the sensitivity of determining $\gamma$
to $y$ and $\epsilon$ for small values of $r_f$.

We now consider the time-dependent decay rate into $f_DK_S$ for a state
$B^0(t)$ which has evolved from an initially pure $B^0$ state.
[A similar expression applies to $\Gamma(B_s(t)\to f_D\phi)$].
Using our previously defined notations~(\ref{eq:defs}) and the
standard notations
\beq \label{defs}
\tau\equiv \Gamma t~,\quad
x\equiv\frac{\Delta m}{\Gamma}~,\quad
\lambda_f\equiv \frac{q}{p}\frac{\bar A_{Bf}}{A_{Bf}}~,
\eeq
where a $B$ flavor index ($d$ or $s$) is implicit,
one obtains~\cite{Bigi:1987in},
\beqa
  \Gamma\lt[B^0(t)\to f_DK_S\rt] &=&
   \frac{e^{-\tau}}{2}\lt|A_{Bf}\rt|^2
   \lt[(\cosh y\tau+\cos x\tau)+|\lambda_f|^2
   (\cosh y\tau-\cos x\tau)\right.\nonumber\\
   &-& 2(\Re\,\lambda_f)\sinh y\tau+2(\Im\,\lambda_f)\sin x\tau\left.\rt]~.
   \label{eq:Brate}
\eeqa
Similarly, for an initial $\bar B^0$ one has
\beqa
\Gamma\lt[\bar B^0(t)\to f_DK_S\rt]
   &=& \frac{e^{-\tau}}{2}\lt|A_{Bf}\rt|^2|p/q|^2
   \lt[(\cosh y\tau-\cos x\tau)+|\lambda_f|^2
   (\cosh y\tau+\cos x\tau)\right.\nonumber\\
   &-& 2(\Re\,\lambda_f)\sinh y\tau-2(\Im\,\lambda_f)\sin x\tau\left.\rt]~.
   \label{eq:barBrate}
\eeqa
The untagged decay rate for a final $f_DK_S$ state is given by
\beqa
\Gamma\lt[B^0(t)\to f_DK_S\rt] &+&
 \Gamma\lt[\bar B^0(t)\to f_DK_S\rt]  =\nonumber\\
 & &  \frac{e^{-\tau}}{2}\Af2\lt\{
   \lt(1+|p/q|^2\rt)\,\lt[\lt(1 + \lt|\lambda_f\rt|^2\rt)\cosh y\tau
  - 2(\Re\,\lambda_f)\,\sinh y\tau\rt] \rt. \nonumber\\
  &+& \lt. \lt(1-|p/q|^2\rt)\,\lt[\lt(1 - \lt|\lambda_f\rt|^2\rt)\cos x\tau
  + 2(\Im\,\lambda_f)\,\sin x\tau\rt]\rt\}~.
 \label{eq:rates}
\eeqa

The total number of decays into $f_DK_S$ is obtained by integrating
over time.
To leading order in $y$ and $\epsilon$, the time-integrated rate is
\beqa \label{master}
   \Gamma_f
&=& \Af2+\Abf2-2y\lt|A_{Bf}\bar A_{Bf}\rt|\cos\varphi_f\nonumber\\
   &-&\epsilon\lt[\frac{x}{x^2+1}\lt|A_{Bf}\bar A_{Bf}\rt|\sin\varphi_f
   -\frac{x^2}{2\lt(x^2+1\rt)}\lt(\Af2-\Abf2\rt)\rt]\nonumber\\
   &+&\Ord\lt(y^2\rt)+\Ord\lt(\epsilon^2\rt)~,  \label{eq:Nf}
\eeqa
where we have defined $\varphi_f\equiv \arg\lambda_f$.
Similarly, for the CP-conjugate final state $\bar f_DK_S$ one has
\beqa
   \Gamma_{\bar f}
&=& \Afb2+\Abfb2-2y\lt|A_{B\bar f}\bar A_{B\bar f}
   \rt|\cos\varphi_{\bar f}\nonumber\\
   &-&\epsilon\lt[\frac{x}{x^2+1}\lt|A_{B\bar f}\bar A_{B\bar f}
   \rt|\sin\varphi_{\bar f}
   -\frac{x^2}{2\lt(x^2+1\rt)}\lt(\Afb2-\Abfb2\rt)\rt]\nonumber\\
   &+&\Ord\lt(y^2\rt)+\Ord\lt(\epsilon^2\rt)~. \label{eq:Nfbar}
\eeqa
Note that this result applies regardless of whether the neutral $B$ ($B_s$) mesons 
are produced hadronically or  in $e^+e^-$ collisions at $\Upsilon(4S)$ 
[$\Upsilon(5S)$].  In the latter case one neutral $B$ ($B_s$) meson is observed decaying 
into $f_DK_S$ ($f_D\phi$), summing over all the decay modes of the second 
$B$ ($B_s$).

As expected,  Eqs.~(\ref{eq:Nf}) and \eqref{eq:Nfbar} reproduce the result of \cite{GGSSZ}
in the limit $y\to 0, \epsilon \to 0$.
We see that the entire $x$-dependence is suppressed by $\epsilon$.
In the case of $B_s$ decays, where $x\gg 1$, we may expand in $1/x$,
obtaining
\beqa
   \Gamma_f
&=& \Af2+\Abf2
   -2y\lt|A_{Bf}\bar A_{Bf}\rt|\cos\varphi_f
   -\frac{\epsilon}{2}\lt(\Af2-\Abf2\rt)\nonumber\\
   &+&\Ord\lt(y^2\rt)+\Ord(\epsilon/x)+\Ord\lt(\epsilon^2\rt)~.
   \label{eq:Nf-Bs}
\eeqa

Note that while the sum of the first two terms in Eqs.~(\ref{eq:Nf}) and \eqref{eq:Nfbar}
depends on two combinations of $B$ decay parameters, 
$A_n^2(1 + r^2_n)$ and $A_n^2r_n\cos\delta_n$, the terms proportional to $y$ 
and $\epsilon$ involve $A_n, r_n$ and $\delta_n$ as three separate variables.
This will affect the counting of parameters required for determining $\gamma$
which we discuss next.

It was shown in \cite{GGSSZ} that, when neglecting $y$ and $\epsilon$, $\gamma$ can
be determined from untagged $B^0$ decays using several final states $f_D^k$.
This holds true also when including nonzero values for $y$ and $\epsilon$,
where the number of unknowns is increased by one, as mentioned above.  
First, assume that $y$ and $\epsilon$ are known.
Consider $N$ different final states, $f_D^k\quad (k=1,\ldots,N)$ which are not 
CP-eigenstates (e.g. $f_D^k = K^-\pi^+, K^{*+}K^-$), and their corresponding
CP-conjugates, $\bar f_D^k$. 
Assume that $A_f^k, A_{\bar f}^k$ and $r_f^k, r_{\bar f}^k$ have been determined 
through branching ratio measurements in an independent sample of $D^0$ decays.
For $N\ge 4$, the $N+4$ unknown parameters $A_n,r_n,\delta_n,\delta_f^k$, and $\gamma$ can 
then be extracted by solving the $2N$ equations (\ref{eq:Nf}) and (\ref{eq:Nfbar}).  
(As a cross check for the solution one may use isospin in order 
to obtain estimates for $A_n$ and $r_n$~\cite{GGSSZ}.)
Including $B^0\to D^*K_S$ followed by $D^*\to D^0\pi^0, D^*\to D^0\gamma$, 
where the two strong phases $\delta_n^*$ differ by $\pi$~\cite{Bondar:2004bi},
one has $6N$ measurements and $N+7$ variables, a system which is solvable for 
$N\ge 2$.

In principle, one may measure in this way not only $\gamma$ but also $y$ and $\epsilon$.
Including these two variables as unknowns permits determining $\gamma$, for $N\ge 6$
when using only the ground state $D$ meson, and for 
$N\ge 2$ when using also decays involving $D^*$. In practice, the values of $y$ 
and $\epsilon$ may be too small
for a useful determination. The procedure above is expected to suffer from a large 
statistical error, and should therefore be used together with studying charged $B$ decays.
As we show in the next section, neglecting $y$ and $\epsilon$
simplifies the equations for neutral $B$ decays, but introduces a systematic error which 
can be quite large for flavor-specific $D$ decays.

\section{The induced theoretical error}
\label{discussion}

Let us now consider the error in $\gamma$ made by assuming $y=\epsilon=0$, for
the special case when $D$ mesons decay to flavor states for which $r_f\ll 1$.
 In order to estimate the sensitivity of $\gamma$ to small nonzero
values of $y$ and $\epsilon$, we will take all other parameters as given
(in the previous section we have outlined a procedure for obtaining
a global solution for all parameters).
Assuming that one measures the rate $\Gamma_f$ for a certain decay mode,
Eq.~(\ref{eq:Nf}) leads to a constraint of the form
$\Gamma_f^{\rm exp}=\Gamma_f(\gamma;y,\epsilon)$.

The dependence of $\Gamma_f$ on $\gamma$
in the limit $y=\epsilon =0$ is given
by the first two terms on the right-hand-side of Eq.~(\ref{eq:Nf}),
\beq\label{eqn:A^2}
\Af2+\Abf2 \propto  1+ r^2_n + 4r_nr_f\cos\delta_n\cos(\delta_f + \gamma) +\Ord(r_f^2)~,
\eeq
where a proportionality factor $A^2_nA^2_f$ was omitted. The effect of nonzero $y$ and $\epsilon$ is given by the remaining terms in Eq.~(\ref{eq:Nf}).
The terms in (\ref{eq:Nf}) which are most sensitive to
$y$ and $\epsilon$ are of zeroth order in $r_f$,
\beq\label{eqn:y}
-2y\lt|A_{Bf}\bar A_{Bf}\rt|\cos\varphi_f \propto
-2yr_n\,\cos(\delta_n-2\beta -\gamma) + \Ord(r_f)~,
\eeq
\beq\label{eqn:epsilon}
\frac{\epsilon x}{x^2 +1}\lt|A_{Bf}\bar A_{Bf}\rt|\sin\varphi_f
\propto \frac{\epsilon x}{x^2 +1}r_n\,\sin(\delta_n-2\beta -\gamma) +\Ord(r_f)~.
\eeq
Thus, while in (\ref{eqn:A^2}) the term which depends on $\gamma$ is linear in $r_f$,
the $\gamma$-dependent terms in (\ref{eqn:y}) and (\ref{eqn:epsilon})
do not involve this small ratio. This implies that the sensitivity of determining $\gamma$ to
$y$ and $\epsilon$ is enhanced by $1/r_f$. The corrections to $\gamma$
from $y\ne 0$ and $\epsilon \ne 0$ are given by terms of the form
\beq
\Delta\gamma^{y}
\sim \frac{y}{r_f}~, \qquad
\Delta\gamma^{\epsilon}
\sim \frac{\epsilon}{r_f}\frac{x}{x^2 +1}~.
\eeq
so that the errors on $\gamma$ may be significant for $f_D=K^+\pi^-$ where $r_{K^+\pi^-} 
\simeq 0.06$~\cite{pdg}.

Using a recent evaluation of mixing parameters for the $B_s$ 
system~\cite{Lenz:2006hd}, $y_s=0.064\pm0.012$,
\mbox{$\epsilon_s=(4.2\pm1.2)\cdot 10^{-5}$}, and the measurement
\cite{HFAG} $x_s=26 \pm 1$,
one notes that $\Delta\gamma^{\epsilon}$ can be safely neglected,
while $\Delta\gamma^{y}$ can be $\Ord(1)$ in $B_s$ decays.
Keeping the $y$-dependence when extracting $\gamma$ from time-integrated
untagged $B_s\to D\phi, D\eta^{(')}$ decays is thus mandatory.
On the other hand, corrections in the $B_d$ system, where one has measured
\cite{HFAG} 
$x_d=0.776\pm0.008$, and one estimates~\cite{Lenz:2006hd}
$y_d=(2.1\pm0.5)\cdot 10^{-3}$, $\epsilon_d=(-9.6\pm2.2)\cdot 10^{-4}$,
are at a level of several percent.
In this case one needs to consider corrections from $y,\epsilon\ne0$
when data reach high statistics.

We stress that the above analysis provides only a rule
of thumb for the size of the expected error in $\gamma$.
The actual error in the analysis will depend crucially on the choice of
the final states, and should be re-evaluated on a case by case basis.
Large corrections from nonzero
$y$ and $\epsilon$ apply only to flavor states in $D$ decays,
where $r_f$ is small. Adding information from singly Cabibbo-suppressed 
$D$ decays where $r_f\sim \Ord(1)$ is expected to dilute the error.

\section{Multi Body $D$ Decays}
\label{threebody}

Next we move to the case of multi-body $D$ decays, such as $D\to K_S\pi^+\pi^-$. 
The error introduced by neglecting $y$ and $\epsilon$ may
be enhanced by quasi two-body decays into flavor states occurring in some parts of 
phase space, for instance in $\bar D^0\to K^{*+}\pi^-$.
We first consider the model-independent approach
presented in \cite{GGSSZ,GGSZ}. 
We find that keeping $y$ and
$\epsilon$ finite makes this approach non-practical, while the corrections can be implemented
if the $D$ decay amplitude is modeled by a sum of Breit-Wigner forms.
We will focus on the three-body decay $D\to K_S\pi^-\pi^+$; however
the following discussion is rather generic.

Neglecting CP violation in $D$ decays, one has
\beqa
   A\lt(\bar D^0\to K_S(p_1),\pi^-(p_2),\pi^+(p_3)\rt)
   &=& A\lt(D^0\to K_S(p_1),\pi^-(p_3),\pi^+(p_2)\rt)\nonumber\\
   &\equiv& A(s_{13},s_{12})\,e^{i\delta(s_{13},s_{12})}~,
\eeqa
where $s_{ij}\equiv(p_i+p_j)^2$. Denoting
\beqa
   A_{ij}\equiv A(s_{1i},s_{1j}),\qquad
   \delta_{ij}\equiv \delta(s_{1i},s_{1j})~,
\eeqa
the amplitude of the cascade decay becomes
\beqa
   A_{Bf} \equiv A(B^0\to f_D K)
    = A_n\lt[A_{32}e^{i\delta_{32}}
    + r_n\,e^{(i\delta_n-\gamma)}\,A_{23}
    \,e^{i\delta_{23}}\rt]~,
\eeqa
and therefore
\beq
   \lt|A_{Bf}\rt|
   =A_n\sqrt{A_{32}^2+r_n^2A_{23}^2+2A_{32}A_{23}r_n\,
   \cos(\gamma+\delta_{23}-\delta_{32}+\delta_n)}~.
   \label{eq:ABf-multi}
\eeq
Similarly,
\beq
   \lt|\bar A_{Bf}\rt|
   =A_n\sqrt{A_{23}^2+r_n^2A_{32}^2+2A_{23}A_{32}r_n\,
   \cos(\gamma+\delta_{23}-\delta_{32}-\delta_n)}~.
   \label{eq:AbBf-multi}
\eeq
These are the analogues of Eq.~(\ref{eq:abs}).

Using these magnitudes, the differential untagged decay rates,
$d^2\Gamma_f/ds_{12}\,ds_{13}$ and 
$d^2\Gamma_{\bar f}/ds_{12}\,ds_{13}$,
are given by Eq.~(\ref{eq:rates}).  Dividing the Dalitz plot
domain $(s_{12},s_{13})$ into bins, the time-integrated numbers of
events over the $i$th bin are given by integrating Eq.~(\ref{eq:Nf})
over this bin. Thus, we have for the $i$th bin,
\beq
   N_i = A_n^2\lt(
   \Sigma^+_i+\frac{\epsilon}{2}\frac{x^2}{x^2+1}
   \Sigma^-_i-2y\Lambda^c_i+\frac{\epsilon x}{x^2+1}\Lambda^s_i\rt)
   + \Ord\lt(y^2\rt)+\Ord\lt(\epsilon^2\rt)~,
     \label{eq:N3}
\eeq
where
\beqa
   \Sigma^+_i &\equiv& (1+r_n^2)T^+_i
   +4r_n\cos\delta_n(c_i\cos\gamma-s_i\sin\gamma)~,  \nonumber\\
   \Sigma^-_i &\equiv& (1-r_n^2)T^-_i
   -4r_n\sin\delta_n(c_i\sin\gamma+s_i\cos\gamma)~,  \nonumber\\
   \Lambda^c_i &\equiv&
   \int_ids_{12}\,ds_{13}\,\lt|A_{Bf}\bar A_{Bf}\rt|\cos\varphi~,
   \nonumber\\
   \Lambda^s_i &\equiv&
   \int_ids_{12}\,ds_{13}\,\lt|A_{Bf}\bar A_{Bf}\rt|\sin\varphi~,
\eeqa
and
\beqa
   T^\pm_i &\equiv& \int_ids_{12}\,ds_{13}\,\lt(A_{32}^2\pm A_{23}^2\rt)~,
   \nonumber\\
   c_i &\equiv& \int_ids_{12}\,ds_{13}\,
   A_{23}A_{32}\cos\lt(\delta_{23}-\delta_{32}\rt)~,   \nonumber\\
   s_i &\equiv& \int_ids_{12}\,ds_{13}\,
   A_{23}A_{32}\sin\lt(\delta_{23}-\delta_{32}\rt)~.
\eeqa

In order to see how many bins are needed if $x,y$ and $\epsilon$
were known, we consider $m$ $B$-decay modes and divide the
$D$-decay Dalitz plot into $k$ bins.
For each $B$ decay mode one has three unknowns,
$A_n,r_n,\delta_n$.
Assuming that the amplitudes $T^\pm_i$ are known from $D$ 
decay data implies four unknowns per bin, 
$c_i,s_i,\Lambda^c_i,\Lambda^s_i$.
If the bins are CP symmetric, the number of independent unknowns
$c_i,s_i$ reduces by half \cite{GGSZ}, while no such symmetry
reduction applies to $\Lambda^c_i,\Lambda^s_i$.
Using both $B^0\to DK_S$ and $B^0\to D^*K_S$, the number of unknowns 
including $\gamma$ is $3k + 7$, while the number of measurements is
$3k$. Such a system of equations is unsolvable.
In principle, one could obtain sufficient information for the purpose of using known 
$y$ and $\epsilon$ to extract $\gamma$ from neutral $B$ decays involving 
$D\to K_S\pi^+\pi^-$ alone. This would require including higher $D$ resonances, 
or multi-body $B$ decays~\cite{multi}. However, this information and precise knowledge of 
$y$ and $\epsilon$ are not expected to be experimentally accessible very soon.

We conclude that in the cases where one cannot neglect $y$ and $\epsilon$ (see the 
discussion in section~\ref{discussion}), the model-independent analysis of untagged
multi-body $D$-decays becomes impractical.  A Breit-Wigner
modeling of the density of events in the Dalitz plot~\cite{GGSZ,DalitzBabar,DalitzBelle},
could be done if $y$ and $\epsilon$ were known.
Without this a priori knowledge, the error in $\gamma$ 
could be large in regions of phase space where doubly Cabibbo-suppressed 
decays are important. A detailed study of the error would be needed.

\section{Summary}
\label{summary}
We have shown that the method presented in~\cite{GGSSZ} for extracting the
CKM phase $\gamma$ from untagged $B_d$ decays leads to a systematic error
caused by neglecting $y$ and $\epsilon$. We found that the terms proportional
to $y$ and $\epsilon$ are enhanced by a ratio of large to small
$D$ decay amplitudes. This enhancement is particularly
notable when using doubly Cabibbo-suppressed $D$ decays. 
It can lead to an error on $\gamma$ at the level of several percent in $B^0\to D K_S$ 
decays, and to an uncertainty of order one when extracting  $\gamma$ in
untagged $B_s\to D\phi, D\eta^{(')}$. 

The method proposed in Ref.~\cite{GGSSZ} can 
thus be implemented without change in present $B\to D^{(*)}K^{(*)}$ data sets. 
However, future precision determinations of $\gamma$ will require taking into account 
corrections from $y$ and $\epsilon$.  
While in $B_s$ decays $\epsilon_s$ can be safely 
set to zero, the effect of  a nonzero $y_s$ cannot be neglected, and experimental 
information of this parameter must be used already at the current  level of precision. 
In the future, the discussed effects may have an impact on a global fit to $\gamma$
including both charged and neutral $B$ decays.  
We also note that at least in principle, with
unlimited statistics, a measurement of $y$ and $\epsilon$ may be performed
by using only time-independent experimental data.

\acknowledgments
This work is supported in part by the Israel Science Foundation
under Grants No. 378/05 and 1052/04 and by by the German-Israeli 
Foundation under Grant No.\ I-781-55.14/2003. The work of JZ was supported in part by the Slovenian Ministry of Science and Education.




\end{document}